\documentclass[aps,twocolumn,prd,showpacs,floatfix]{revtex4}
\usepackage{bm}
\usepackage{amsmath}
\usepackage{nicefrac}
\usepackage{amsthm}
\usepackage{amssymb}
\usepackage{graphicx}

\newcommand{\etal}{{\it et al.}}
\newcommand{\Eref}[1]{Eq.~(\ref{#1})}

\newcommand{\appropto}{\mathrel{\vcenter{
  \offinterlineskip\halign{\hfil$##$\cr
    \propto\cr\noalign{\kern2pt}\sim\cr\noalign{\kern-2pt}}}}}

\begin{document}

\title{Enhanced effects of temporal variation of the fundamental constants
in $^{2}\Pi_{1/2}$ term diatomic molecules: $^{207}$Pb$^{19}$F}

\date{\today}

\author{V.~V.~Flambaum$^{1}$}
\author{Y.~V.~Stadnik$^{1}$}
\author{M.~G.~Kozlov$^{1,2,3}$}
\author{A.~N.~Petrov$^{2,4}$}

\affiliation{$^{1}$School of Physics, University of New South Wales, Sydney 2052,
Australia}
\affiliation{$^{2}$Petersburg Nuclear Physics Institute, Gatchina, 188300, Russia}
\affiliation{$^{3}$St.~Petersburg Electrotechnical University ``LETI'', Prof. Popov Str. 5, 197376
St.~Petersburg, Russia}
\affiliation{$^{4}$Division of Quantum Mechanics, St.~Petersburg State University, 198904, Russia}

\begin{abstract}
The $^{207}$Pb$^{19}$F molecule possesses a pair of closely spaced levels of
opposite parity due to near cancelation of the omega-type doubling and
magnetic hyperfine interaction energy shifts [Alphei \etal\ Phys. Rev. A,
\textbf{83}, 040501 (2011)]. We calculate the dependence of the transition
frequency between these levels  on the fine-structure constant $\alpha$ and
the ratio of the light quark masses to the quantum chromodynamics scale
$\left(m_{q}/\Lambda_{\mathrm{QCD}}\right)$, and find large enhancement of the
relative effects of the variation of these parameters.  Note that the effect
of $\alpha$ variation appears mainly due to the significant difference in the
relativistic correction factors for the fine and hyperfine structure. We hence
suggest the $^{207}$Pb$^{19}$F molecule as a candidate system for
investigating the possible temporal variation of the fundamental constants.

\end{abstract}


\maketitle

\section{Introduction}
\label{Sec:Intro}

The $^{207}$Pb$^{19}$F molecule has been recently studied as a system with
high sensitivity to P-odd and P,T-odd interactions \cite{Alphei11,MMB11}.
Here we want to point out that this molecule is also very sensitive to the
variation of the fundamental constants of nature. The idea that such constants
might vary with time can be traced as far back as the Large Numbers Hypothesis
of Dirac, who hypothesized that the gravitational constant $G$ might be
proportional to the reciprocal of the age of the universe
\cite{Dirac37,Dirac38,Dirac74}. In more recent times, the possibility of
observing the space-time variation of the fundamental constants of nature has
received renewed interest, with the possible variation of the fine-structure
constant $\alpha=e^2/\hbar c$ and the electron-to-proton mass ratio $m_e/m_p$
receiving the most attention (see e.g. \cite{Webb,Kozlov,Murphy,Bagdonaite}). Another dimensionless fundamental parameter for
strong interactions is $m_q/\Lambda_\mathrm{QCD}$, where $m_q$ is the
light quark mass and $\Lambda_\mathrm{QCD}$ is the QCD scale
\cite{FlSh02,Flambaum09}. This parameter enters atomic physics through nuclear
magnetic $g$-factors \cite{Flambaum03,Flambaum04,Tedesco06}.

At present, there are very strong upper bounds on the time variation of these
constants from laboratory experiments
(see e.g. \cite{Peik04,Fortier07,Blatt08,Rosenband08,Shelk08,Guena12,Budker07,Budker13}).
More details about the current status of these experiments and on
observational astrophysical and geophysical data can be found in a number of
reviews \cite{Peik10,Chi11,Uza11,BeFl11,BFO13}. All experiments cited above
were performed with atoms, the only exception being experiment \cite{Shelk08},
which was done on the ro-vibrational transition of the SF$_6$ molecule. However,
recently there has been a growing number of proposals to use different types of
molecular transitions, where sensitivity to the variation of the fundamental
constants is strongly enhanced compared to typical atomic transitions
\cite{Flambaum06,FK07b,DSS08,ZKY08}. Molecules are also playing a very important
role in astronomical studies of the possible variation of the fundamental
constants on the cosmological time-scale (see, for example, review
\cite{KoLe13}).

In the present work, we propose another system for testing the possible
temporal variation of the fundamental constants. Closely spaced levels of
opposite parity are known to exist in the $^{207}$Pb$^{19}$F molecular
radical species, with the separation between the levels experimentally known
to be $\omega = 266.285$ MHz \cite{Alphei11}. The close spacing between these
levels is a result of the near cancelation between the shifts in the energies
of these levels due to omega-type doubling and the magnetic hyperfine
interaction. We show that the dependences on $\alpha$ and
$m_{q}/\Lambda_{\mathrm{QCD}}$ of these two nearly cancelling
contributions are significantly different for omega-type doubling and magnetic
hyperfine shifts, resulting in a sizeable enhancement in the $\alpha$- and
$\left(m_{q}/\Lambda_{\mathrm{QCD}}\right)$-dependence of the transition
frequency $\omega$. Note that the effect of $\alpha$ variation appears mainly due to
the large difference in the relativistic correction factors for the fine and
hyperfine structure. The molecular energy levels of interest are also quite
stable, making experiments using the $^{207}$Pb$^{19}$F molecule attractive
for tests of the possible temporal variation of the fundamental constants.

The structure of this paper is as follows. In Section \ref{Sec:non-lin}, we derive the non-linear dependence of the transition frequency $\omega$ on the omega-doubling, rotational and hyperfine interaction parameters. In Sec.~\ref{Sec:alpha}, we
derive the dependence of the transition frequency $\omega$ on $\alpha$. To
solve this problem, we estimate higher order corrections in $Z \alpha$ to the
fine structure and omega doubling intervals.  In Sec.~\ref{Sec:mu}, we derive
the dependence of the transition frequency $\omega$ on
$m_{q}/\Lambda_{\mathrm{QCD}}$. Finally, in Sec.~\ref{Sec:sum}, we present the
formula for the dependence of $\omega$ on both $\alpha$ and
$m_{q}/\Lambda_{\mathrm{QCD}}$, together with estimates for the sensitivity
coefficients, and summarise our findings.

\section{Dependence of transition frequency on omega-doubling and hyperfine interaction parameters}
\label{Sec:non-lin}

The full Hamiltonian for the $^{207}$Pb$^{19}$F molecule is given in
\cite{Alphei11} and can be solved to give the value of $\omega$, provided that
one knows all the necessary values of the rotational and hyperfine structure
constants, see, for instance, Refs.~\cite{Kozlov87,Titov10,Titov13}. The energy levels of interest are the closely spaced $F^{p}=1/2^{+}$ and $F^{p}=1/2^{-}$ states. Noting that the two most dominant contributions to $\omega$ are from omega-type doubling and the
magnetic hyperfine interaction for lead, we ignore the magnetic hyperfine contribution from fluorine (which can be treated as a perturbation) and so instead consider the $F^{p}=1^{+}$ and $F^{p}=1^{-}$ states, the total angular momenta quantum numbers of which arise from the coupling of the angular momentum $J=1/2$ with the nuclear spin angular momentum of lead $I=1/2$ alone. The energies of the $F^{p}=1^{+}$ and $F^{p}=1^{-}$ states can be presented as follows \cite{Ray08}
\begin{equation}
\label{energies}
E\left( J,F,p \right) = BJ(J+1) + p(-1)^{J+1/2} \frac{\Delta}{2}\left(J+\frac{1}{2}\right) + U_{\textrm{hf}} ,
\end{equation}
where
\begin{align}
\label{}
&U_{\textrm{hf}} = \frac{\chi A_{\perp}}{4} \notag \\
&+ 2(J-F) \left[-\tau + s \sqrt{\tau^2 - \beta \left(\frac{\tau}{F+1/2} - \beta \right) } \right] ,
\end{align}
\begin{equation}
\label{}
\chi = (-1)^{F+1} p ,
\end{equation}
\begin{equation}
\label{}
\tau = \left(B+\frac{\chi \Delta}{2} \right)(F+1/2) ,
\end{equation}
\begin{equation}
\label{}
\beta = \frac{A_{\parallel}-\chi A_{\perp}}{4} ,
\end{equation}
\begin{equation}
\label{sign}
s = \textrm{sign} \left[\tau - \frac{\beta}{2F+1} \right] .
\end{equation}
Here $p$ denotes the parity of the state. The values of the rotational constant $B=6917.9108$ MHz, the omega-doubling parameter $\Delta = -4145.2304$ MHz and the hyperfine parameters for lead $A_{\perp}=-7264.0388$ MHz, $A_{\parallel}=10146.6733$ MHz are known experimentally \cite{Alphei11}. In Eq.~(\ref{energies}), the first term represents the pure rotational energy contribution, the second term represents the omega-doubling contribution that is present even in the absence of the hyperfine interaction, while the third term represents the contribution from the hyperfine interaction, which also includes non-linear rotational and omega-doubling contributions. Note that in Ref.~\cite{Ray08}, there is an error in the phase assignment in the molecular wavefunction used to calculate the energy levels. The relative phase between the basis functions of opposite projections of total electronic angular momentum on the molecular axis, linear combinations of which result in molecular wavefunctions of definite parity $p$, should be $(-1)^{J-S}p$. For the $F=0$ states, this results in the substitution $\chi \to -\chi$ in Eqs.~(1) - (4) in Ref.~\cite{Ray08}. For further details, we refer the reader to Refs.~\cite{Titov10,Carrington03}.

We write the energy separation between the two $F=1$ levels of interest in the following form
\begin{equation}
\label{omega}
\omega = E\left( 1/2,1,+ \right) - E\left( 1/2,1,- \right) .
\end{equation}
Variation of (\ref{omega}) with respect to the omega-doubling, hyperfine interaction and rotational parameters leads to
\begin{equation}
\label{delta_omega_new1}
\delta \omega = \frac{\partial \omega}{\partial\Delta} \delta\Delta + \frac{\partial \omega}{\partial A_{\perp}} \delta A_{\perp} + \frac{\partial \omega}{\partial A_{\parallel}} \delta A_{\parallel} + \frac{\partial \omega}{\partial B} \delta B .
\end{equation}
From Eqs.~(\ref{energies}) - (\ref{omega}), we find
\begin{equation}
\label{deriv1}
\frac{\partial \omega}{\partial\Delta} = -0.862 ,
\end{equation}
\begin{equation}
\label{deriv2}
\frac{\partial \omega}{\partial A_{\perp}} = 0.497 ,
\end{equation}
\begin{equation}
\label{deriv3}
\frac{\partial \omega}{\partial A_{\parallel}} = -0.139 ,
\end{equation}
\begin{equation}
\label{deriv4}
\frac{\partial \omega}{\partial B} = 0.272 .
\end{equation}
For comparison, we find from solving the full Hamiltonian (including the hyperfine interaction for fluorine) numerically, the corresponding values of the derivatives in Eqs.~(\ref{deriv1}) - (\ref{deriv4}) to be $-0.863$, $0.493$, $-0.140$ and $0.271$ respectively. In the present work, we use the analytical values given by (\ref{deriv1}) - (\ref{deriv4}) for the derivative values, but for the energy separation we use the experimentally determined value $\omega = 266.285$ MHz \cite{Alphei11}.


\section{Variation of transition frequency with $\alpha$}
\label{Sec:alpha}

It is known that the $\alpha$-dependence of the magnetic hyperfine interaction
energy shift scales as $\alpha^2 F_{\mathrm{rel}}^{\mathrm{hf}} \left(Z \alpha
\right)$, where $F_{\mathrm{rel}}^{\mathrm{hf}} \left(Z \alpha \right)$ is the
Casimir relativistic correction factor, which for s- and p-waves with $j=1/2$
is given approximately by
\begin{equation}
\label{F_rel_hf} F_{\mathrm{rel}}^{\mathrm{hf}} = \frac{3}{\gamma_{1/2}
\left(4 \gamma^2_{1/2} -1 \right)} ,
\end{equation}
see e.g. Refs.~\cite{Prestage95,Khrip91}. We employ the standard notation
$\gamma_{j} = \sqrt{\left(j+1/2\right)^2-\left(Z\alpha\right)^2}$ in
Eq.~(\ref{F_rel_hf}) and throughout this work. According to our numerical
estimate based on the PbF wave function presented in \cite{Kozlov87}, the
magnetic hyperfine structure is dominated by the $p_{1/2}$-wave contribution;
other waves contribute a few per cent only (see also \cite{Schwartz55}).
Variation of $A$, where $A$ is either $A_{\parallel}$ or $A_{\perp}$, with respect to $\alpha$ thus leads to the following
expression
\begin{equation}
\label{delta_F_rel_hf} \frac{\delta A}{A} =
\left(2+K_{\mathrm{rel}}^{\mathrm{hf}}\right) \frac{\delta \alpha}{\alpha} ,
\end{equation}
where $K_{\mathrm{rel}}^{\mathrm{hf}}$ is given by
\begin{equation}
\label{K_rel_hf} K_{\mathrm{rel}}^{\mathrm{hf}} = \frac{\left(Z
\alpha\right)^2 \left(12 \gamma^2_{1/2} - 1 \right)}{\gamma^2_{1/2}
\left(4\gamma^2_{1/2} -1 \right)}
\end{equation}
for both $s_{1/2}$- and $p_{1/2}$-waves. For lead $\left(Z=82\right)$,
Eq.~(\ref{K_rel_hf}) gives $K_{\mathrm{rel}}^{\mathrm{hf}}=2.39$. We note that
more accurate numerical many-body calculations of the dependence of the
hyperfine structure energy shift on $\alpha$ give slightly larger values of
the coefficient $K_{\mathrm{rel}}^{\mathrm{hf}}$ than the analytical Casimir
correction factor does for moderately heavy atomic and ionic species
\cite{Dzuba99}. For instance, for Cs ($Z=55$),
$K_{\mathrm{rel}}^{\mathrm{hf}}=0.83$ numerically (instead of 0.74
analytically), while for Hg$^{+}$ ($Z=80$),
$K_{\mathrm{rel}}^{\mathrm{hf}}=2.28$ numerically (instead of 2.18
analytically). However, for the purposes of the present work, it will suffice
to use the analytical expression (\ref{K_rel_hf}).

The omega-type doubling of interest in the $^{207}$Pb$^{19}$F molecule occurs
between the positive and negative-parity $^{2}\Pi_{1/2}$ states. The Coriolis
interaction can connect the $\Omega=+1/2$ and $\Omega=-1/2$ states to first
order in perturbation theory, but cannot connect the $\Lambda=+1$ and
$\Lambda=-1$ states directly without there being mixing of the $\Lambda=+1$
state with the $\Lambda=0$ state via the spin-orbit interaction, and similarly
mixing of the $\Lambda=-1$ state with the $\Lambda=0$ state (see e.g.~\cite{Kozlov09}). If we consider, for instance, the subspace spanned by the
unperturbed states $\left|\Lambda=+1\right>$ and $\left|\Lambda=0\right>$ in
the two-level approximation, then in the presence of the spin-orbit
interaction between these two states, the perturbed eigenfunction
corresponding to the unperturbed state $\left|\Lambda=+1\right>$ in the lowest
order approximation reads
\begin{equation}
\label{pert} \left|\widetilde{\Lambda=+1}\right> = \xi \left|\Lambda=+1\right>
+ \eta \left|\Lambda=0\right> ,
\end{equation}
where $\left|\eta\right| \ll 1$ is the spin-orbit mixing coefficient. These
considerations imply that the $\alpha$-dependence of the omega-type doubling
energy shift between the positive and negative-parity $^{2}\Pi_{1/2}$ states
in $^{207}$Pb$^{19}$F scales in the same way as $\left|\eta\right|$ does. We
also know that
 \begin{align}
 &\left|\eta\right| \approx
 \left|V_{\mathrm{so}}\right| / \Delta \varepsilon\,,
 \\ \label{DeltaEps}
 &
 \Delta \varepsilon
 \equiv
 E\left({}^{2}\Sigma_{1/2}\right) - E\left({}^{2}\Pi_{1/2}\right)\,,
 \end{align}
which is to say that the $\alpha$-dependence of the omega-type doubling energy
shift scales approximately in the same way as the spin-orbit interaction
energy shift (in atomic units) does. Here $V_{\mathrm{so}}$ denotes the matrix
element of the spin-orbit interaction operator between the states $^{2}
\Sigma_{1/2}$ and $^{2} \Pi_{1/2}$, with projections of the orbital angular
momentum on the molecular symmetry axis being $\Lambda=0$ and $\Lambda=+1$
respectively.

It is well known that the spin-orbit matrix elements in atomic units scale as
$Z^2 \alpha^2$. However, this expression is valid only for small values of
$Z^2 \alpha^2$. For the hyperfine interaction, higher order $Z^2 \alpha^2$
corrections are very important; they even produce a singularity for $Z^2
\alpha^2=3/4$ for the point-like nucleus case - see Eq.~(\ref{F_rel_hf}).
Therefore, we should estimate higher order $Z^2 \alpha^2$ corrections for the
spin-orbit splitting and omega doubling. Relativistic effects, such as the
spin-orbit interaction, arise predominantly at small distances $\left(r
\lesssim a_{B}/Z\right)$, where screening of the nuclear Coulomb field is
negligible \cite{Casimir}. Also, the binding energy of the unpaired electron
in $^{207}$Pb$^{19}$F is small compared with the Coulomb potential energy.
Thus for $r \lesssim a_{B}/Z$, the wavefunction is proportional to the
hydrogen-like ion wavefunction  with a large principal quantum number $n$.
These points imply that the $\alpha$ dependence of the spin-oribit matrix
elements may be found from the spin-orbit splitting $np_{3/2}-np_{1/2}$ in
hydrogen-like ions.
The relativistic Dirac formula for the energy levels of a hydrogen-like
species reads (see e.g. \cite{Sakurai})
\begin{equation}
\label{Dirac_energies_unscaled}
E_{n,j} = \frac{mc^2}{\left[1+\frac{\left(Z\alpha \right)^2}
{\left(\gamma_j+n'\right)^2}  \right]^{1/2}} ,
\end{equation}
where $n=j+1/2+n'$. To work with dimensionless quantities,
we  take the ratio of (\ref{Dirac_energies_unscaled}) to the non-relativistic
energy scale, $mc^2 \left(Z \alpha\right)^2/2n^2$, giving
\begin{equation}
\label{Dirac_energies}
\varepsilon_{n,j} = \frac{2 n ^2} {\left(Z\alpha \right)^2
\left[1+\frac{\left(Z\alpha \right)^2} {\left(\gamma_j+n'\right)^2}
\right]^{1/2}} .
\end{equation}
We take the limit $n' \to \infty$ (note again that $n=j+1/2+n'$) and find that the
$\alpha$-dependence of the energy difference $\varepsilon_{n,3/2} -
\varepsilon_{n,1/2}$ scales as
\begin{equation}
\label{F_rel_omega1} C_{\mathrm{rel}}^\mathrm{so} \equiv \frac{Z^2
\alpha^2}{4}  F_{\mathrm{rel}}^\mathrm{so} = \gamma_{3/2}-\gamma_{1/2}-1 .
\end{equation}
The first term in the expansion of the right-hand-side gives the usual $ Z^2
\alpha^2$ dependence, while higher orders give the relativistic correction
factor $F_{\mathrm{rel}}^\mathrm{so}$ for the  spin-orbit interaction. The
difference between the relativistic correction factor in
Eq.~(\ref{F_rel_omega1}) compared with the relativistic correction factor for
the hyperfine interaction in Eq.~(\ref{F_rel_hf}) is very significant: the
relativistic factor for the spin-orbit interaction remains finite ($<3$) for
any $Z \alpha <1$, i.e. the relativistic corrections are significantly smaller
than that for the hyperfine interaction (which become infinite for $Z^2
\alpha^2=3/4$). Variation of (\ref{F_rel_omega1}) with respect to $\alpha$
gives
\begin{align}
\label{K_rel_omega}
2+  K_{\mathrm{rel}}^{\mathrm{so}} &= \frac{\partial C_{\mathrm{rel}}^\mathrm{so}}{\partial \alpha}
\frac{\alpha}{C_{\mathrm{rel}}^\mathrm{so}}
= \frac{(Z \alpha)^2 \left(\frac{1}{\gamma_{1/2}}-\frac{1}{\gamma_{3/2}}
\right)}{\gamma_{3/2}-\gamma_{1/2}-1} .
\end{align}
For $Z=82$, Eq.~(\ref{K_rel_omega}) gives
$
2+K_{\mathrm{rel}}^\mathrm{so}=2.42$, i.e.
higher order relativistic corrections increase the result by 20\%. This gives
the following variation of $\Delta$ with respect to $\alpha$
\begin{equation}
\label{delta_F_omega} \frac{\delta \Delta}{\Delta} = (2+ K_{\mathrm{rel}}^\mathrm{so})
\frac{\delta \alpha}{\alpha} .
\end{equation}

In this work, we deal with a heavy Pb atom. The spin-orbit interaction rapidly
increases with $Z^2 \alpha^2$. In the hypothetical case of a very large
spin-orbit interaction, which exceeds an interval between the $\Lambda=1$ and
$\Lambda=0$ terms, omega doubling does not depend on the spin-obit
interaction, since the state with definite electronic angular momentum $j=1/2$
already contains both $\Lambda=1$ and $\Lambda=0$ components. This means that,
in this limit, the $\alpha$ dependence of the omega doubling vanishes. In the
realistic case of the PbF molecule, this does not happen. However, there
exists a further non-linear correction to Eq.~(\ref{delta_F_omega}), which
followed from perturbation theory for closely spaced states. We again restrict
our attention to the subspace spanned by the unperturbed states
$\left|\Lambda=+1\right>$ and $\left|\Lambda=0\right>$ in the two-level
approximation. The spin-orbit mixing coefficient in Eq.~(\ref{pert}) can be
expressed as follows \cite{LL3}
\begin{equation}
\label{} \left|\eta\right| =\frac{1}{\sqrt{2}}\sqrt{1-\frac{\Delta
\varepsilon}{\varepsilon}} ,
\end{equation}
where $\Delta\varepsilon$ is defined by \Eref{DeltaEps} and
\begin{equation}
\label{} \varepsilon =\sqrt{\left|\Delta
\varepsilon\right|^2+4\left|V_{\mathrm{so}}\right|^2} .
\end{equation}
Noting that $\left|V_{\mathrm{so}}\right| / \Delta \varepsilon \ll 1$, we find
\begin{equation}
\label{b_approx} \left|\eta \right| \approx
\frac{\left|V_{\mathrm{so}}\right|}{\left|\Delta \varepsilon \right|} \left[1-
\frac32\frac{\left|V_{\mathrm{so}}\right|^2}{\left|\Delta \varepsilon
\right|^2} \right] ,
\end{equation}
from which the following equation follows directly
\begin{align}
\label{b_approx'} \frac{\partial\left|\eta\right|}{\partial \alpha}
\frac{\alpha}{\left|\eta\right|} &=
\frac{\partial\left|V_{\mathrm{so}}\right|}{\partial \alpha}
\frac{\alpha}{\left|V_{\mathrm{so}}\right|} \left[\frac{1-
\frac92\left|V_{\mathrm{so}}\right|^2/\left|\Delta \varepsilon \right|^2}
{1-\frac32 \left|V_{\mathrm{so}}\right|^2/ \left|\Delta \varepsilon
\right|^2}\right]
\notag \\
&\approx (2+ K_{\mathrm{rel}}^\mathrm{so}) \left[
{1- 3\left|V_{\mathrm{so}}\right|^2/ \left|\Delta \varepsilon \right|^2 }
\right]
.
\end{align}
The second term in square brackets in the last line of Eq.~(\ref{b_approx'})
is the non-linear correction factor $\chi$ to Eq.~(\ref{delta_F_omega})
\begin{equation}
\label{delta_F_omega_corr} \frac{\delta \Delta}{\Delta} =  (2+
K_{\mathrm{rel}}^\mathrm{so}) \chi \frac{\delta \alpha}{\alpha} .
\end{equation}
From numerical calculations \cite{Kozlov87}, we know that
$\left|V_{\mathrm{so}} / \Delta \varepsilon\right| = 0.19$ and so we find that
$\chi=0.89$.

Finally, we estimate the $\alpha$-dependence of the contribution of the
molecular rotational constant $B$  to the variation of the energy separation
$\omega$ as follows. The effect of this contribution to the sensitivity
coefficient for $\alpha$ in our final expression~(\ref{delta_w-w-exp-pred}) is
very small, so a refined calculation is not necessary here. The rotational
constant of interest here ($B\equiv B_{1/2}$) is that for the $^{2}\Pi_{1/2}$
state, which near the Pb nucleus is dominated by the $p_{1/2}$ atomic orbital
(over 80\% -- see \cite{Kozlov87}).
The $\alpha$-dependence of $B_{1/2}$ arises due to the relativistic correction
to the potential $V^{\textrm{rel}}(r)$, which is located near the Pb nucleus
where the Coulomb potential is not screened and the energy of the valence
electron may be neglected. In this region, the valence electron wave function is
proportional to the Coulomb wave function with a large principal quantum
number $n$. From the relativistic energy shifts of the high Coulomb levels in
Eq.~(\ref{Dirac_energies_unscaled}), we know that
$\langle np_{1/2}|V^{\textrm{rel}}|np_{1/2}\rangle \approx
2\langle np_{3/2}|V^{\textrm{rel}}|np_{3/2}\rangle$. This gives the following relations between the
relativistic shifts of the rotational constants
\begin{equation}\label{B}
B_{1/2}^{\textrm{rel}}=2 B_{3/2}^{\textrm{rel}}=2 ( B_{1/2}-B_{3/2})=\mathrm{const} \left(Z \alpha\right)^2 \,,
\end{equation}
where $B_{3/2}$ is the rotational constant for the $^{2}\Pi_{3/2}$ state, and from which we find that
\begin{equation}
\label{}
\frac{\delta B_{1/2}}{B_{1/2}} = \frac{2 B_{1/2}^{\textrm{rel}}}{B_{1/2}} \frac{\delta \alpha}{\alpha} \equiv 2 \nu \frac{\delta \alpha}{\alpha} .
\end{equation}
We use the experimentally determined rotational constant values $B_{3/2} = 0.23403$ cm$^{-1}$ and $B_{1/2} = 0.22875$ cm$^{-1}$ for $^{208}$Pb$^{19}$F \cite{NIST}, giving $\nu = -0.046$.


\section{Variation of transition frequency with $m_{q}/\Lambda_{\mathrm{QCD}}$}
\label{Sec:mu} The omega-type doubling and rotational energy shifts are obviously independent
of the nuclear magnetic moment $\mu$. The $\mu$-dependence of the hyperfine
interaction energy shift, however, scales linearly with $\mu$. The variation
of $A$ with respect to $\mu_{\mathrm{Pb}}$ is thus
\begin{equation}
\label{delta_A_mu} \frac{\delta A}{A} = \frac{\delta
\mu_{\mathrm{Pb}}}{\mu_{\mathrm{Pb}}} ,
\end{equation}
which can also be expressed as follows \cite{Tedesco06}
\begin{equation}
\label{delta_A_X} \frac{\delta A}{A} = \kappa_{\mathrm{Pb}} \frac{\delta
\left(m_{q}/\Lambda_{\mathrm{QCD}}\right)}{\left(m_{q}/\Lambda_{\mathrm{QCD}}\right)}
.
\end{equation}
Noting that there is little sensitivity of $\frac{\delta \mu}{\mu}$ to core
polarization effects in odd-neutron, even-proton nuclei \cite{Tedesco06}, we
can estimate $\kappa_{\mathrm{Pb}}$ from known data. The $^{207}$Pb nucleus in
the ground state has $I^{\pi}=1/2^{-}$ and nuclear magnetic moment
$\mu=+0.592583~\mu_{N}$, while the $^{199}$Hg nucleus in the ground state has
$I^{\pi}=1/2^{-}$ and nuclear magnetic moment $\mu=+0.5058855~\mu_{N}$
\cite{Data05}. Without account of nuclear radius variation,
$\kappa_{\mathrm{Hg}}=-0.09$ \cite{Tedesco06,Kava11}. However, with account of
the effect of nuclear radius variation on the hyperfine structure,
$\kappa_{\mathrm{Hg}}=-0.111$ \cite{Dinh09}. Since the values and origin of
the $^{207}$Pb and $^{199}$Hg nuclear magnetic moments, as well as their radii
are similar, we take $\kappa_{\mathrm{Pb}} \approx
\kappa_{\mathrm{Hg}}=-0.111$ for our estimate of $\kappa_{\mathrm{Pb}}$.

\section{Summary and Conclusions}
\label{Sec:sum} The variation of $\omega$ (in atomic units) with respect to
$\alpha$ and $m_{q}/\Lambda_{\mathrm{QCD}}$ reads
\begin{align}
\label{delta_w-total}
\frac{\delta \omega}{\omega} &= \left[ \frac{ \chi (2+K_{\mathrm{rel}}^\mathrm{so})}{\omega} \Delta \frac{\partial \omega}{\partial \Delta} + \frac{2\nu}{\omega} B \frac{\partial \omega}{\partial B}  \right. \notag \\
&\left.+ \frac{\left(2+K_{\mathrm{rel}}^{\mathrm{hf}}\right)}{\omega} \left( A_{\parallel} \frac{\partial \omega}{\partial A_{\parallel}} + A_{\perp} \frac{\partial \omega}{\partial A_{\perp}} \right) \right] \frac{\delta
\alpha}{\alpha}  \notag \\
&+ \frac{\kappa_{\mathrm{Pb}}}{\omega} \left( A_{\parallel} \frac{\partial \omega}{\partial A_{\parallel}} + A_{\perp} \frac{\partial \omega}{\partial A_{\perp}} \right)  \frac{\delta
\left(m_{q}/\Lambda_{\mathrm{QCD}}\right)} {\left(m_{q}/\Lambda_{\mathrm{QCD}}\right)} .
\end{align}
Substituting all the known quantities into Eq.~(\ref{delta_w-total}) and taking into account that $\Delta$, $A_{\perp}$, $A_{\parallel}$ and $B$ all have linear dependence on the electron-to-proton mass ratio, $m_e/m_p$, gives
\begin{equation}
\label{delta_w-w-exp-pred} \frac{\delta \omega}{\omega} \approx -55
\frac{\delta \alpha}{\alpha} + 2.1  \frac{\delta
\left(m_{q}/\Lambda_{\mathrm{QCD}}\right)}{\left(m_{q}/\Lambda_{\mathrm{QCD}}\right)} +\frac{\delta \left(m_e/m_p \right)}{\left(m_e/m_p\right)}
.
\end{equation}
With the approximations made in deriving relation (\ref{delta_w-w-exp-pred}),
the uncertainties in the sensitivity coefficients in
(\ref{delta_w-w-exp-pred}) are $\sim 20 \%$. Note that the effect of the variation of $m_e/m_p$ is not enhanced.

We see that the $^{207}$Pb$^{19}$F molecular radical species can offer a one
to two order of magnitude enhancement of the relative effect of
$\alpha$-variation. This is comparable to the enhancements in some other
molecular species \cite{Flambaum06,FK07b,DSS08,ZKY08}. Even more interestingly, the
sensitivity coefficient for $m_{q}/\Lambda_{\mathrm{QCD}}$ is enhanced by two
orders of magnitude compared with the ratio of frequencies of $^{133}$Cs and
$^{87}$Rb atomic clocks, which use electronic hyperfine transitions as their
frequency standards \cite{Tedesco06,Dinh09,Kava11} and currently provide the
best limit on the variation of $m_{q}/\Lambda_{\mathrm{QCD}}$ \cite{Guena12}.
Furthermore, the natural widths of the closely spaced energy levels of
interest are quite small, since both states lie merely $\sim 8000$ MHz above
the ground state. An additional advantage is that this molecule is already
considered for high precision experiments to study the P-odd anapole moment of
the nucleus $^{207}$Pb and to search for the electron EDM \cite{Alphei11}. We
hence suggest the $^{207}$Pb$^{19}$F molecule as a candidate system for
investigating the possible temporal variation of the fundamental constants.

Since the effects of the variation of fundamental constants in the $^{207}$Pb$^{19}$F molecule are significantly enhanced, it does not matter
what system will be used to provide the reference frequency. In fact, we
presented the result of the variation of the ratio $\omega/$(atomic unit). In
order to use a specific frequency standard, such as caesium or rubidium, one
should subtract the effect of the variation of the corresponding standard
frequency, which is presented (also in atomic units) in
Refs.~\cite{Tedesco06,Dinh09,Kava11}. This will provide only small corrections
to Eq.~(\ref{delta_w-w-exp-pred}).

\section*{ACKNOWLEDGEMENTS}
We are grateful to Anatoly V.~Titov for important discussions. This work is supported by the Australian Research Council. Alexander N.~Petrov would like to acknowledge support from the SPbU Fundamental Science Research grant from Federal budget No.~0.38.652.2013 and RFBR Grant No.~13-02-01406.




\begin{thebibliography}{99}
\expandafter\ifx\csname natexlab\endcsname\relax\def\natexlab#1{#1}\fi
\expandafter\ifx\csname bibnamefont\endcsname\relax
  \def\bibnamefont#1{#1}\fi
\expandafter\ifx\csname bibfnamefont\endcsname\relax
  \def\bibfnamefont#1{#1}\fi
\expandafter\ifx\csname citenamefont\endcsname\relax
  \def\citenamefont#1{#1}\fi
\expandafter\ifx\csname url\endcsname\relax
  \def\url#1{\texttt{#1}}\fi
\expandafter\ifx\csname urlprefix\endcsname\relax\def\urlprefix{URL }\fi
\providecommand{\bibinfo}[2]{#2} \providecommand{\eprint}[2][]{\url{#2}}

\bibitem{Alphei11} L. D. Alphei,
J.-U. Grabow, A. N. Petrov, R. Mawhorter, B. Murphy, A. Baum, T. J. Sears, T.
Zh. Yang, P. M. Rupasinghe, C. P. McRaven, and N. E. Shafer-Ray, Phys. Rev. A
\textbf{83}, 040501(R), (2011).


\bibitem[{\citenamefont{Mawhorter et~al.}(2011)\citenamefont{Mawhorter, Murphy,
  Baum, Sears, Yang, Rupasinghe, {McRaven}, {Shafer-Ray}, Alphei, and
  Grabow}}]{MMB11}
\bibinfo{author}{\bibfnamefont{R.~J.} \bibnamefont{Mawhorter}},
  \bibinfo{author}{\bibfnamefont{B.~S.} \bibnamefont{Murphy}},
  \bibinfo{author}{\bibfnamefont{A.~L.} \bibnamefont{Baum}},
  \bibinfo{author}{\bibfnamefont{T.~J.} \bibnamefont{Sears}},
  \bibinfo{author}{\bibfnamefont{T.}~\bibnamefont{Yang}},
  \bibinfo{author}{\bibfnamefont{P.~M.} \bibnamefont{Rupasinghe}},
  \bibinfo{author}{\bibfnamefont{C.~P.} \bibnamefont{{McRaven}}},
  \bibinfo{author}{\bibfnamefont{N.~E.} \bibnamefont{{Shafer-Ray}}},
  \bibinfo{author}{\bibfnamefont{L.~D.} \bibnamefont{Alphei}},
  \bibnamefont{and} \bibinfo{author}{\bibfnamefont{J.-U.}
  \bibnamefont{Grabow}}, \bibinfo{journal}{Phys. Rev. A}
  \textbf{\bibinfo{volume}{84}}, \bibinfo{pages}{022508}
  (\bibinfo{year}{2011}).


\bibitem{Dirac37} P.~A.~M.~Dirac, Nature \textbf{139}, 323, (1937).

\bibitem{Dirac38} P.~A.~M.~Dirac, Proc.~R.~Soc.~Lond.~A \textbf{165}, 921, (1938).

\bibitem{Dirac74} P.~A.~M.~Dirac, Proc.~R.~Soc.~Lond.~A \textbf{338}, 439, (1974).

\bibitem{Webb} J.~K.~Webb, J.~A.~King, M.~T.~Murphy, V.~V.~Flambaum, R.~F.~Carswell, and M.~B.~Bainbridge, Phys.~Rev.~Lett. \textbf{107}, 191101 (2011); arXiv:1008.3907.

\bibitem{Kozlov} V.~V.~Flambaum, and M.~G.~Kozlov, Phys.~Rev.~Lett., \textbf{98}, 240801
(2007); arXiv:0704.2301.

\bibitem{Murphy} M.~T.~Murphy, V.~V.~Flambaum, S.~Muller, and
 C.~Henkel, Science, \textbf{320}, 1611 (2008); arXiv:0806.3081.

\bibitem{Bagdonaite} J.~Bagdonaite, P.~Jansen, C.~Henkel, H.~L.~Bethlem, K.~M.~Menten, and W.~Ubachs, Science, \textbf{339}, 46 (2013).

\bibitem[{\citenamefont{{Flambaum} and {Shuryak}}(2002)}]{FlSh02}
\bibinfo{author}{\bibfnamefont{V.~V.} \bibnamefont{{Flambaum}}}
  \bibnamefont{and} \bibinfo{author}{\bibfnamefont{E.~V.}
  \bibnamefont{{Shuryak}}}, \bibinfo{journal}{Phys. Rev. D}
  \textbf{\bibinfo{volume}{65}}, \bibinfo{eid}{103503} (\bibinfo{year}{2002});
  \eprint{arXiv:hep-ph/0201303}.

\bibitem{Flambaum09} V.~V.~Flambaum, and R.~B.~Wiringa, Phys.~Rev.~C \textbf{79}, 034302, (2009).

\bibitem{Flambaum03} V.~V.~Flambaum; arXiv:physics/0309107.

\bibitem{Flambaum04} V.~V.~Flambaum, D.~B.~Leinweber, A.~W.~Thomas, and R.~D.~Young, Phys.~Rev.~D \textbf{69}, 115006 (2004); arXiv:hep-ph/0402098.

\bibitem{Tedesco06} V.~V.~Flambaum, and A.~F.~Tedesco, Phys.~Rev.~C \textbf{73}, 055501, (2006).










\bibitem{Peik04} E.~Peik, B.~Lipphardt, H.~Schnatz, T.~Schneider, Chr.~Tamm, and S.~G.~Karshenboim, Phys.~Rev.~Lett. \textbf{93}, 170801, (2004).


\bibitem{Fortier07} T.~M.~Fortier \textit{et al}., Phys.~Rev.~Lett. \textbf{98}, 070801, (2007).

\bibitem{Blatt08} S.~Blatt \textit{et al}., Phys.~Rev.~Lett. \textbf{100}, 140801, (2008).

\bibitem{Rosenband08} T.~Rosenband \textit{et al}., Science \textbf{319}, 1808, (2008).

\bibitem{Shelk08} A.~Shelkovnikov, R.~J.~Butcher, C.~Chardonnet, and A.~Amy-Klein, Phys.~Rev.~Lett. \textbf{100}, 150801, (2008).

\bibitem{Guena12} J.~Gu$\acute{\mathrm{e}}$na, M.~Abgrall, D.~Rovera, P.~Rosenbusch, M.~E.~Tobar, P.~Laurent, A.~Clairon, and S.~Bize, Phys.~Rev.~Lett. \textbf{109}, 080801, (2012).

\bibitem{Budker07}  S.~J.~Ferrell, A.~Cing\"oz, A.~Lapierre, A.-T.~Nguyen, N.~Leefer,
D.~Budker, V.~V.~Flambaum, S.~K.~Lamoreaux, and J.~R.~Torgerson, Phys.~Rev.~A \textbf{76}, 062104 (2007); arXiv:0708.0569.


\bibitem{Budker13} N.~Leefer, C.~T.~M.~Weber, A.~Cing\"oz, J.~R.~Torgerson, and D.~Budker, Phys.~Rev.~Lett. \textbf{111}, 060801, (2013).


\bibitem{Peik10} E.~Peik, Nuc.~Phys.~B (Proc.~Suppl.) \textbf{203-204}, 18, (2010).




\bibitem[{\citenamefont{{Chiba}}(2011)}]{Chi11}
\bibinfo{author}{\bibfnamefont{T.}~\bibnamefont{{Chiba}}},
  \bibinfo{journal}{Progress of Theoretical Physics}
  \textbf{\bibinfo{volume}{126}}, \bibinfo{pages}{993} (\bibinfo{year}{2011});
  \eprint{arXiv:1111.0092}.

\bibitem[{\citenamefont{Uzan}(2011)}]{Uza11}
\bibinfo{author}{\bibfnamefont{J.-P.} \bibnamefont{Uzan}},
  \bibinfo{journal}{Living Reviews in Relativity}
  \textbf{\bibinfo{volume}{14}}, \bibinfo{pages}{2} (\bibinfo{year}{2011}),
  \urlprefix\url{http://www.livingreviews.org/lrr-2011-2}.

\bibitem[{\citenamefont{{Berengut} and {Flambaum}}(2011)}]{BeFl11}
\bibinfo{author}{\bibfnamefont{J.~C.} \bibnamefont{{Berengut}}}
  \bibnamefont{and} \bibinfo{author}{\bibfnamefont{V.~V.}
  \bibnamefont{{Flambaum}}}, \bibinfo{journal}{Journal of Physics Conference
  Series} \textbf{\bibinfo{volume}{264}}, \bibinfo{eid}{012010}
  (\bibinfo{year}{2011}); \eprint{arXiv:1009.3693}.

\bibitem[{\citenamefont{{Berengut} et~al.}(2013)\citenamefont{{Berengut},
  {Flambaum}, and {Ong}}}]{BFO13}
\bibinfo{author}{\bibfnamefont{J.~C.} \bibnamefont{{Berengut}}},
  \bibinfo{author}{\bibfnamefont{V.~V.} \bibnamefont{{Flambaum}}},
  \bibnamefont{and} \bibinfo{author}{\bibfnamefont{A.}~\bibnamefont{{Ong}}},
 in  \emph{\bibinfo{booktitle}{European Physical Journal Web of Conferences}}
  (\bibinfo{year}{2013}), vol.~\bibinfo{volume}{57}
 of \emph{\bibinfo{series}{European Physical Journal Web of Conferences}},
   p. \bibinfo{pages}{2001}.






\bibitem{Flambaum06} V.~V.~Flambaum, Phys.~Rev.~A \textbf{73}, 034101, (2006).


\bibitem[{\citenamefont{Flambaum and Kozlov}(2007)}]{FK07b}
\bibinfo{author}{\bibfnamefont{V.~V.} \bibnamefont{Flambaum}}, \bibnamefont{and}
  \bibinfo{author}{\bibfnamefont{M.~G.} \bibnamefont{Kozlov}},
  \bibinfo{journal}{Phys. Rev. Lett.} \textbf{\bibinfo{volume}{99}},
  \bibinfo{pages}{150801} (\bibinfo{year}{2007}); \eprint{arXiv:0705.0849}.

\bibitem[{\citenamefont{DeMille et~al.}(2008)\citenamefont{DeMille, Sainis,
  Sage, Bergeman, Kotochigova, and Tiesinga}}]{DSS08}
\bibinfo{author}{\bibfnamefont{D.}~\bibnamefont{DeMille}},
  \bibinfo{author}{\bibfnamefont{S.}~\bibnamefont{Sainis}},
  \bibinfo{author}{\bibfnamefont{J.}~\bibnamefont{Sage}},
  \bibinfo{author}{\bibfnamefont{T.}~\bibnamefont{Bergeman}},
  \bibinfo{author}{\bibfnamefont{S.}~\bibnamefont{Kotochigova}},
  \bibnamefont{and} \bibinfo{author}{\bibfnamefont{E.}~\bibnamefont{Tiesinga}},
  \bibinfo{journal}{Phys. Rev. Lett.} \textbf{\bibinfo{volume}{100}},
  \bibinfo{pages}{043202} (\bibinfo{year}{2008});
  \bibinfo{note}{arXiv:\eprint{0709.0963}}.

\bibitem[{\citenamefont{Zelevinsky et~al.}(2008)\citenamefont{Zelevinsky,
  Kotochigova, and Ye}}]{ZKY08}
\bibinfo{author}{\bibfnamefont{T.}~\bibnamefont{Zelevinsky}},
  \bibinfo{author}{\bibfnamefont{S.}~\bibnamefont{Kotochigova}},
  \bibnamefont{and} \bibinfo{author}{\bibfnamefont{J.}~\bibnamefont{Ye}},
  \bibinfo{journal}{Phys. Rev. Lett.} \textbf{\bibinfo{volume}{100}},
  \bibinfo{pages}{043201} (\bibinfo{year}{2008});
  \bibinfo{note}{arXiv:\eprint{0708.1806}}.

\bibitem[{\citenamefont{Kozlov and Levshakov}(2013)}]{KoLe13}
\bibinfo{author}{\bibfnamefont{M.~G.} \bibnamefont{Kozlov}} \bibnamefont{and}
  \bibinfo{author}{\bibfnamefont{S.~A.} \bibnamefont{Levshakov}},
  \bibinfo{journal}{Annalen der Physik} \textbf{\bibinfo{volume}{525}},
  \bibinfo{pages}{452} (\bibinfo{year}{2013}); \bibinfo{note}{arXiv:1304.4510}.






\bibitem{Kozlov87} M.~G.~Kozlov, V.~I.~Fomichev, Yu.~Yu.~Dmitriev, L.~N.~Labzovsky, and A.~V.~Titov, J.~Phys.~B \textbf{20}, 4939, (1987).

\bibitem{Titov10} K.~I.~Baklanov, A.~N.~Petrov, A.~V.~Titov, and M.~G.~Kozlov, Phys.~Rev.~A \textbf{82}, 060501(R), (2010).

\bibitem{Titov13} A.~N.~Petrov, L.~V.~Skripnikov, A.~V.~Titov, and R.~J.~Mawhorter, Phys.~Rev.~A \textbf{88}, 010501(R), (2013).

\bibitem{Ray08} C.~P.~McRaven, P.~Sivakumar, and N.~E.~Shafer-Ray, Phys.~Rev.~A \textbf{78}, 054502, (2008).

\bibitem{Carrington03} J.~M.~Brown, and A.~Carrington, \emph{Rotational Spectroscopy of Diatomic Molecules}, (Cambridge University Press, Cambridge, 2003).


\bibitem{Prestage95} J.~D.~Prestage, R.~L.~Tjoelker, and L.~Maleki, Phys.~Rev.~Lett. \textbf{74}, 3511, (1995).

\bibitem{Khrip91} I.~B.~Khriplovich, \emph{Parity Nonconservation in Atomic Phenomena}, (Gordon and Breach, Philadelphia, 1991).

\bibitem{Schwartz55} C.~Schwartz, Phys.~Rev. \textbf{97}, 380, (1955).

\bibitem{Dzuba99} V.~A.~Dzuba, V.~V.~Flambaum, and J.~K.~Webb, Phys.~Rev.~A \textbf{59}, 230, (1999).

\bibitem{Kozlov09} M.~G.~Kozlov, Phys.~Rev.~A \textbf{80}, 022118, (2009).

\bibitem{Casimir} H.~B.~G.~Casimir, \emph{On the Interaction Between Atomic Nuclei and Electrons}, (Teyler's Tweede Genootschap, Haarlem, 1936).

\bibitem{Sakurai} J.~J.~Sakurai, and J.~Napolitano, \emph{Modern Quantum Mechanics}, 2nd Ed. (Addison-Wesley, San Fransisco, 2011).

\bibitem{LL3} L.~D.~Landau, and E.~M.~Lifshitz, \emph{Quantum Mechanics (Non-relativistic Theory)}, 3rd Ed. (Butterworth-Heinemann, Oxford, 1977).


\bibitem{NIST} National Institute of Standards and Technology, \emph{Chemistry WebBook}, last updated 2011, \urlprefix\url{http://webbook.nist.gov/}.


\bibitem{Data05} N.~J.~Stone, At.~Data.~Nucl.~Data.~Tables \textbf{90}, 75, (2005).

\bibitem{Kava11} J.~C.~Berengut, V.~V.~Flambaum, and E.~M.~Kava, Phys.~Rev.~A \textbf{84}, 042510, (2011).

\bibitem{Dinh09} T.~H.~Dinh, A.~Dunning, V.~A.~Dzuba, and V.~V.~Flambaum, Phys.~Rev.~A \textbf{79}, 054102, (2009).




\end{thebibliography}
\end{document}